\documentclass[aps,prx,twocolumn,showpacs,psfig,superscriptaddress,longbibliography]{revtex4-2}

\usepackage{times}
\usepackage{graphicx}
\usepackage{float}
\usepackage{latexsym,amsmath,amssymb,bm,euscript}
\usepackage{color}
\usepackage{subfigure}
\usepackage{epstopdf}
\usepackage[colorlinks=true,linkcolor=blue,citecolor=blue]{hyperref}
\usepackage{soul}
\usepackage[normalem]{ulem}
\usepackage{mathrsfs}
\usepackage{amsmath}
\usepackage{lettrine}
\usepackage{bbding}
\usepackage{xspace}
\usepackage{textcomp}
\usepackage{textcase}
\usepackage{setspace}
\usepackage{xcolor}
\usepackage{lineno}

\def\para{\ensuremath{/\kern -0.8em /}\xspace}

\def\beqn{\begin{eqnarray}}
\def\eeqn{\end{eqnarray}}
\def\beq{\begin{equation}}
\def\eeq{\end{equation}}

\newcommand{\Beq}{\begin{eqnarray*} }
\newcommand{\Eeq}{\end{eqnarray*} }
\newcommand{\Bmat}{\left(\begin{matrix}}
\newcommand{\Emat}{\end{matrix}\right)}

\begin{document}

\title{Possible Intermediate Quantum Spin Liquid Phase in  
$\alpha$-RuCl$_3$ under High Magnetic Fields up to 100 T}

\author{Xu-Guang~Zhou}
\thanks{These authors contributed equally to this work.}
\affiliation{Institute for Solid State Physics, University of Tokyo, Kashiwa, Chiba 277-8581, Japan}

\author{Han~Li}
\thanks{These authors contributed equally to this work.}
\affiliation{Kavli Institute for Theoretical Sciences, University of Chinese Academy of Sciences, Beijing 100190, China}
\affiliation{Peng Huanwu Collaborative Center for Research and Education $\&$ School of Physics, Beihang University, Beijing 100191, China}

\author{Yasuhiro~H.~Matsuda}
\email{ymatsuda@issp.u-tokyo.ac.jp}
\affiliation{Institute for Solid State Physics, University of Tokyo, Kashiwa, Chiba 277-8581, 
Japan}

\author{Akira~Matsuo}
\affiliation{Institute for Solid State Physics, University of Tokyo, Kashiwa, Chiba 277-8581, 
Japan}

\author{Wei Li}
\email{w.li@itp.ac.cn}
\affiliation{CAS Key Laboratory of Theoretical Physics, Institute of Theoretical Physics, 
Chinese Academy of Sciences, Beijing 100190, China}
\affiliation{Peng Huanwu Collaborative Center for Research and Education $\&$ School of Physics, Beihang University, Beijing 100191, China}

\author{Nobuyuki Kurita}
\affiliation{Department of Physics, Tokyo Institute of Technology, Tokyo 152-8551, Japan}

\author{Gang Su}
\affiliation{Kavli Institute for Theoretical Sciences, University of Chinese Academy of Sciences, Beijing 100190, China}

\author{Koichi Kindo}
\affiliation{Institute for Solid State Physics, University of Tokyo, Kashiwa, Chiba 277-8581, Japan}

\author{Hidekazu~Tanaka}
\affiliation{Department of Physics, Tokyo Institute of Technology, Tokyo 152-8551, Japan}

\maketitle


\noindent{\bf{Abstract}}\\
\textbf{Pursuing the exotic quantum spin liquid (QSL) state in the 
Kitaev material $\alpha$-RuCl$_3$ has intrigued great research 
interest recently. A fascinating question is on the possible 
existence of a field-induced QSL phase in this compound. Here we 
perform high-field magnetization measurements of $\alpha$-RuCl$_3$ 
up to 102~T employing the non-destructive and destructive pulsed 
magnets. Under the out-of-plane field along the $\textbf{c}^*$ axis (i.e.,
perpendicular to the honeycomb plane), two quantum phase transitions 
are uncovered at respectively 35~T and about 83~T, between which 
lies an intermediate phase as the predicted QSL. This is in sharp 
contrast to the case with in-plane fields, where a single transition 
is found at around 7~T and the intermediate QSL phase is absent 
instead. By measuring the magnetization data with fields tilted 
from the $\textbf{c}^*$ axis up to $90^\circ$ (i.e., in-plane direction), 
we obtain the field-angle phase diagram that contains the zigzag,
paramagnetic, and QSL phases. Based on the $K$-$J$-$\Gamma$-$\Gamma^{\prime}$ 
model of $\alpha$-RuCl$_3$ with 
a large Kitaev term we perform density matrix renormalization 
group simulations and reproduce the quantum phase diagram in excellent 
agreement with experiments.\\}

\noindent{\bf{Introduction}}\\
Quantum spin liquid (QSL) constitutes a topological state of matter 
in frustrated magnets, where the constituent spins remain disordered 
even down to absolute zero temperature and share long-range quantum
entanglement~\cite{Anderson1973,Balents2010,Zhou2017,Broholm2020}. 
Due to the lack of rigorous QSL ground states, such ultra quantum 
spin states are less well-understood in systems in more than one 
spatial dimension before Alexei Kitaev introduced the renowned 
honeycomb model with bond-dependent exchange~\cite{Kiteav2006}. 
The ground state of the Kitaev honeycomb model is proven to be 
a QSL with two types of fractional excitations~\cite{Kiteav2006,
Hermanns2018}. Soon after, the Kitaev model was proposed to be 
materialized in the $J_{\rm eff}=1/2$ Mott insulating magnets
\cite{Trebst2017arXiv,Winter2017,Takagi2019,Janssen2019,Motome2020} 
such as A$_2$IrO$_3$ (A = Li and Na)~\cite{Jackeli2009,Singh2012},
$\alpha$-RuCl$_3$~\cite{Kim2015,KimHS2016}, etc.

Recently, the 4$d$ spin-orbit magnet $\alpha$-RuCl$_3$ has been widely 
accepted as a prime candidate for Kitaev material~\cite{Plumb2014,Sears2015,
Johnson2015,Banerjee2016,Banerjee2017,Ran2017,Sears2020}. As initially 
proposed from the first-principle analysis~\cite{Kim2015,KimHS2016,Winter2016,
Wang2017,Winter2017NC}, the compound is now believed to be described 
by the $K$-$J$-$\Gamma$-$\Gamma^{\prime}$ effective model that includes 
the Heisenberg $J_{(1,3)}$, Kitaev exchange $K$, and the symmetric off-diagonal 
exchange $\Gamma^{(\prime)}$ terms. The Kitaev interaction originates from 
chlorine-mediated exchange through edge-shared octahedra arranged on 
a honeycomb lattice. Similar to the intensively studied honeycomb and 
hyperhoneycomb iridates~\cite{Chaloupka2013}, additional non-Kitaev terms 
$\Gamma^{(\prime)}$ and/or $J_{3}$, unfortunately, stabilize a zigzag 
antiferromagnetic order below $T_N$ $\approx$ 7~K in the compound
\cite{Sears2015,Johnson2015,Banerjee2017,Kubota2015}. Given that, 
a natural approach to realizing the Kitaev QSL is to suppress the zigzag 
order by applying magnetic fields to the compound~\cite{Zheng2017,Baek2017,
Leahy2017torque,Jansa2018,Wulferding2020,Banerjee2018,Balz2019,Wellm2018,
Ponomaryov2020,Hentrich2019Large,Kasahara2018Unusual,Kasahara2018,
Yokoi2021,Czajka2021,Tanaka2022}. As shown in certain experiments, 
a moderate in-plane field (about 7~T) can suppress the zigzag order 
and may induce an intermediate QSL phase before the polarized 
phase under in-plane fields~\cite{Balz2019,Wellm2018,Kasahara2018,
Yokoi2021,Czajka2021}. However, there are also experimental pieces 
of evidence from, e.g., magnetization~\cite{Kubota2015,Johnson2015}, 
magnetocaloric~\cite{Bachus2020}, magneto-torque measurements
\cite{Modic2021}, etc., that indicate a single transition scenario 
with no intermediate phase present. 
Some angle-dependent experiments, on the other hand, demonstrate 
the presence of an additional intermediate phase, which however is, 
due to another zigzag antiferromagnetic order induced by six-layer 
periodicity along the out-of-plane direction~\cite{lampenkelley2021}. 
This leaves an intriguing question to be resolved in the compound 
$\alpha$-RuCl$_3$.

\begin{figure*}[t]
\includegraphics[width = 0.8 \linewidth]{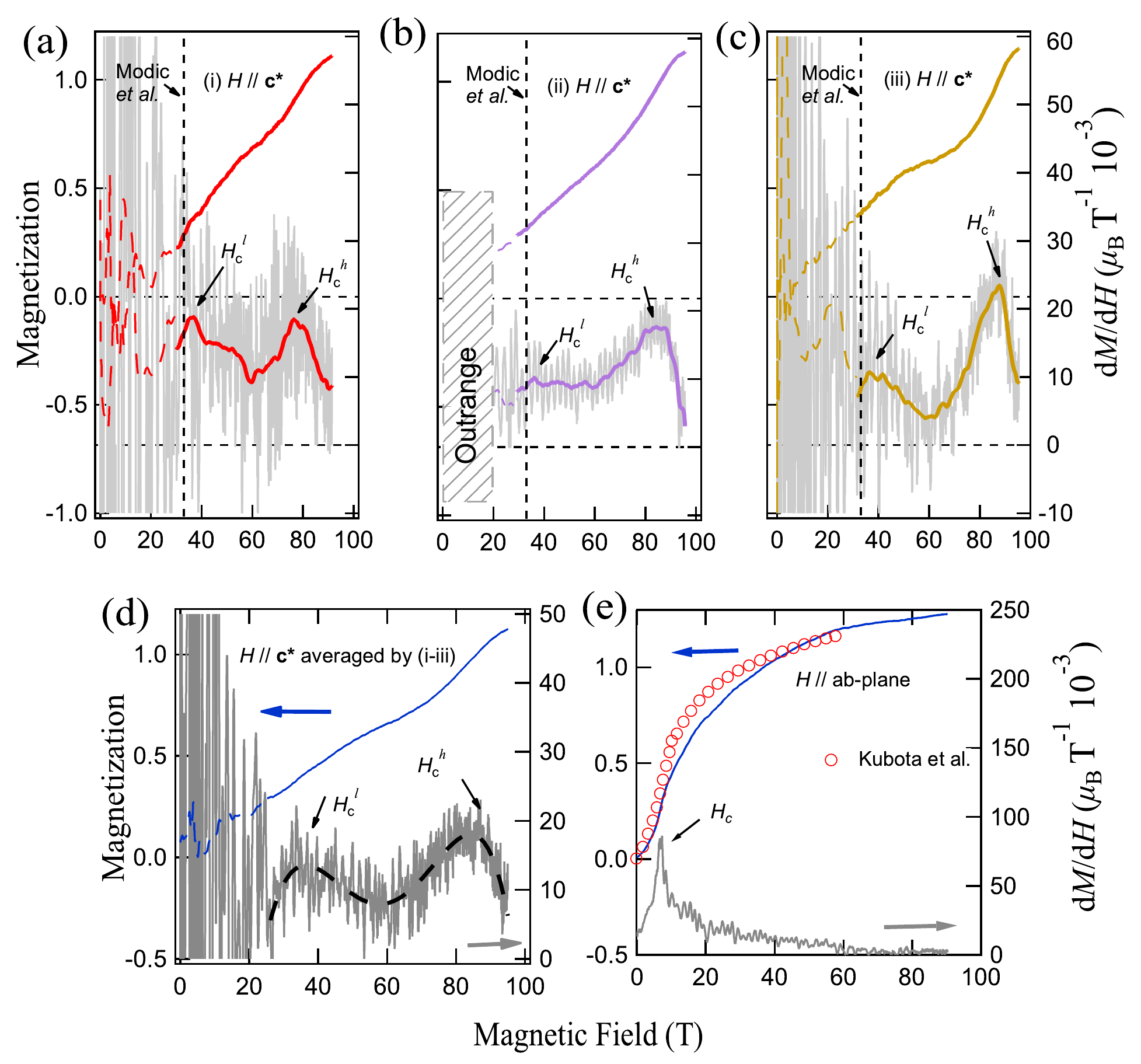}
\caption{\textbf{The magnetization process of $\alpha$-RuCl$_3$ up to 100 T.} 
(a-c) The d$M$/d$H$ data (lower) measured up to 100~T
under out-of-plane fields ($H$ $\parallel$ $\textbf{c}^*$)  and the integrated magnetization 
curves (upper). The grey oscillating curves are the raw d$M$/d$H$ data, with 
the smoothed lines also presented. The data from 0 to 30~T is shown as dash 
line because of the strong starting switch noise~\cite{Zhou2020, Matsuda2013}. 
(i), (ii) and (iii) represent three independent experiments showing similar 
results despite an uncertainty in field angles of $\pm 2.5^\circ$, and experiment 
(ii) is performed with the high-frequency-cut filters. The shadow range 
($\leq$ 20~T) in (ii) is not precisely measured because of the outranged 
noise. The transition field along $\textbf{c}^{*}$ reported by Modic \textit{et al.}
\cite{Modic2021} is also marked by the vertical dashed line. (d) The 
averaged d$M$/d$H$ and $M-H$ data from experiment (i), (ii), and (iii), where the two 
phase-transition signals can be more clearly seen. The black dashed line is 
a guide for the eye. (e) The high-field magnetization measurements under 
in-plane fields up to 90~T, where only a single transition at about $7$~T 
is observed, in excellent agreement with previous measurements by Kubota 
\textit{et al.} (Ref.~\cite{Kubota2015}).}
\label{exp_M}
\end{figure*}

Theoretical progress lately suggests the absence of intermediate QSL
under in-plane fields, while predicting the presence of an intermediate 
phase by switching the magnetic fields from in-plane to the much less 
explored out-of-plane direction. The numerical calculations
\cite{Riedl2019,Gordon2019,Wang2019,Lee2020nc,Janssen2017} of the 
$K$-$J$-$\Gamma$-$\Gamma^{\prime}$ spin model show that the off-diagonal 
exchanges $\Gamma^{(\prime)}$ terms dominate the magnetic anisotropy 
in the compound. Due to the strong magnetic anisotropy in $\alpha$-RuCl$_3$, 
the critical field increases dramatically from the in-plane to the 
out-of-plane direction.
The authors in Ref.~\cite{Gordon2019} further point out an interesting 
two-transition scenario with a field-induced intermediate QSL phase, 
which is later confirmed by other theoretical calculations~\cite{Lee2020nc}, 
except for subtlety in lattice rotational symmetry breaking (such 
a (so-called) nematic order is, however, not directly relevant to our 
experimental discussion here as the realistic compound $\alpha$-RuCl$_3$ 
does not strictly have a $C_3$ symmetry~\cite{KimHS2016,Winter2016,
Leahy2017torque}). More recently, H. Li \textit{et al.} proposed  
a large Kitaev-term spin Hamiltonian~\cite{Lih2021} also based on 
the $K$-$J$-$\Gamma$-$\Gamma^{\prime}$ model. With the precise model 
parameters determined from fitting the experimental thermodynamics 
data, they theoretically reproduced the suppression of zigzag order 
under the 7-T in-plane field, and find a gapless QSL phase located 
between two out-of-plane transition fields that are about 35~T and 
of 100-T class, respectively. Therefore, the previously unsettled 
debates on the field-induced transitions and the concrete theoretical 
proposal of the intermediate QSL phase strongly motivate a high-field 
experimental investigation on $\alpha$-RuCl$_3$ along the out-of-plane
direction and up to 100 T. 

In this work, we report the magnetization ($M$) process of $\alpha$-RuCl$_3$ 
by applying magnetic fields ($H$) in various directions within the honeycomb 
plane and along the $\textbf{c}^*$ axis (out-of-plane) up to 100~T, 
and find clear experimental evidence supporting the two-transition scenario. 
Here, the $\textbf{c}^*$ axis is the axis perpendicular to the honeycomb plane
\cite{Kubota2015}. Under fields applied along and close to the $\textbf{c}^{*}$ axis, 
an intermediate phase is found bounded by two transition fields $H_{c}^{l}$ 
and $H_{c}^{h}$. In particular, besides the previously reported $H_{c}^l 
\simeq 32.5$~T~\cite{Modic2018,Modic2021}, remarkably we find a second phase 
transition at a higher field $H_{c}^{h}\approx83$~T. Below $H_{c}^h$ and 
above $H_{c}^l$ there exists an intermediate phase --- the predicted 
field-induced QSL phase~\cite{Gordon2019,Lih2021}. When the field tilts 
an angle from the $\textbf{c}^*$ axis by 9$^{\circ}$, only the transition field 
$H_c$ is observed, indicating the intermediate QSL phase disappears. 
Accordingly, we also perform the density-matrix renormalization group 
(DMRG) calculations based on the previously proposed 
$K$-$J$-$\Gamma$-$\Gamma^{\prime}$ model of $\alpha$-RuCl$_3$, and find 
the simulated phase transitions and extended QSL phase are in agreement 
with experiments. Therefore, we propose a complete field-angle phase 
diagram and provide the experimental evidence for the field-induced QSL 
phase in the prominent Kitaev compound $\alpha$-RuCl$_3$.\\

\noindent{\bf{Results}}\\
{\textbf{Experimental Results.}}
Figure~\ref{exp_M}~(a-c) shows the magnetization process and the 
magnetic field dependence of d$M$/d$H$ along the $\textbf{c}^*$ (out-of-plane) 
direction. The 
magnetization data represented by the dash lines (0 T to 30~T) are very 
noisy because of the huge switching electromagnetic noise inevitably 
generated for injection mega-ampere driving currents at the beginning 
of the destructive ultra-high field generation~\cite{Zhou2020}. 
The magnetization process and d$M$/d$H$ are precisely measured from 
30 to 95~T, which shows two peaks labeled by $H_{c}^{l}$ and 
$H_{c}^{h}$. To be specific, we have conducted three independent 
measurements (i), (ii), and (iii) in Fig.~\ref{exp_M}, where 
$H_{c}^{l}$ is found to be about 35 T in three measurements
\footnote{We also note that the $\sim$36 T signal was not observed 
in the previous magnetization measurement~\cite{Kubota2015,Johnson2015}, 
it maybe caused by the increasing ABAB stacking fault in $\alpha$-RuCl$_3$}, 
and in agreement with the magneto-torque probe result (32.5 T)~\cite{Modic2021} 
(marked with the vertical dashed line in Fig.~\ref{exp_M}). On the other hand, 
the measured $H_{c}^{h}$ fields are somewhat different in cases (i), (ii), 
and (iii), with values of 76 T, 83 T, and 87 T, respectively. This difference 
can be attributed to the small angle ambiguity ($\pm 2.5^\circ$) in the three
measurements and also to the high sensitivity of the transition field for the 
field angle near the $\textbf{c}^*$ axis of the compound~\cite{Riedl2019}. 
Moreover, we average the d$M$/d$H$ curves from experiments (i-iii), 
show the results in Fig.~\ref{exp_M}(d), and find the averaging process 
has significantly reduced the electrical noise. This allows us to 
identify more clearly the two peaks at $H_c^l$ and $H_c^h$, respectively.
 
\begin{figure}[t]
\includegraphics[width = 0.85 \linewidth]{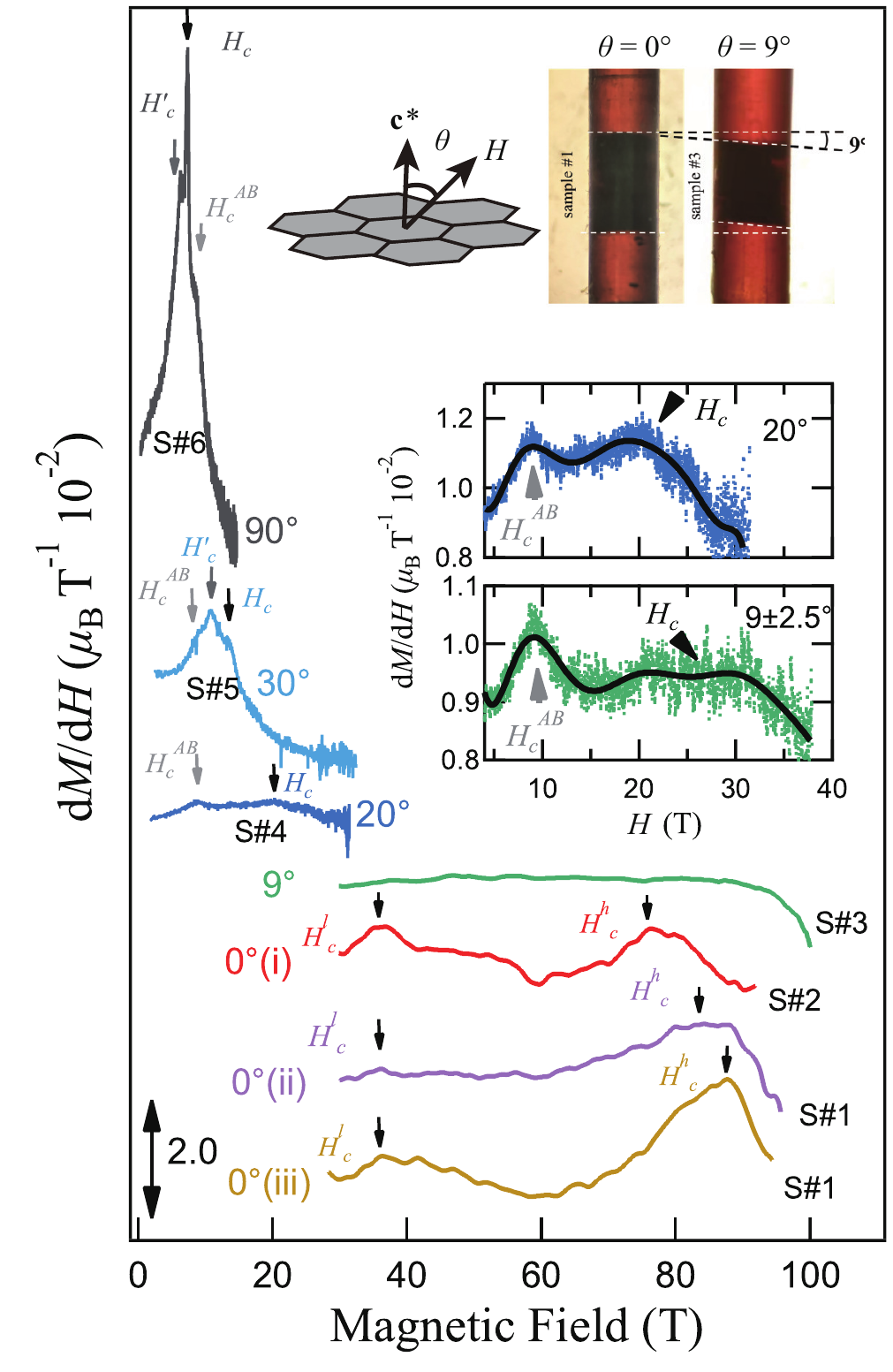}
\caption{\textbf{The d$M$/d$H$ curves at various $\theta$ angles.} We include 
the measurements with $\theta\simeq 0^{\circ}, 9^\circ, 20^\circ, 
30^\circ, 90^\circ$, where the $0^{\circ}$ measurements are performed 
for multiple times (NoS. i, ii, and iii) using the destructive method 
with possible tilting angle within $\pm 2.5^\circ$. Sample \#1-6 represent 
the sample number in different field directions (S\#1-6). The black arrows 
pointing to the peaks of d$M$/d$H$ denote the transition fields in the 
measurements, while the grey ones with $H_c^{\prime}$ and 
$H_c^{AB}$ indicate the irrelevant feature due to the three dimensional 
spin structure~\cite{lampenkelley2021} and the magnetic 
phase transition in sample with ABAB stacking fault, 
respectively. The upper inset illustrates the angle $\theta$ between the 
applied magnetic field and $\textbf{c}^*$ axis, as well as the photos of holding setup 
of the samples for $\theta\simeq 0^{\circ}$ and 9$^\circ$. The two middle insets show the averaged 
d$M$/d$H$ curves obtained by the non-destructive magnet because the 
transition signals are very weak. The black solid curves are guides for 
the eye.
}
\label{exp_dmdh}
\end{figure}

Figure~\ref{exp_dmdh} shows the measured d$M$/d$H$ results for various 
tilting angles ranging from $\theta \simeq 0^\circ$ (i.e., out-of-plane 
fields) to $90^\circ$ (in-plane). For $\theta\simeq 0^\circ$ and $9^\circ$, 
the data are obtained by the destructive method, while the d$M$/d$H$ curves 
with $\theta \simeq 20^{\circ}$, 30$^{\circ}$ and 90$^{\circ}$ are obtained 
by the non-destructive magnet and up to about 30~T.  

The three $\theta\simeq0^\circ$ cases are also plotted in Fig.~\ref{exp_dmdh}. 
Here only the high-quality data above 30~T are shown, which exhibit double 
peaks at $H_{c}^{l}$ and $H_{c}^{h}$. With the single-turn coil technique 
reaching the ultra-high magnetic field of 100~T class, here we are able to 
reach the higher transition field near $H_{c}^{h}\simeq 83$~T that has not 
been reached before. It is noteworthy that although the down-sweep data 
in the field-decreasing measurements are unavailable to be integrated due 
to the field inhomogeneity~\cite{Matsuda2013,Takeyama2012}, nevertheless 
the signals at $H_c^l$ and $H_c^h$ in the up-sweep 
and down-sweep processes are consistent 
(c.f., Supplementary Fig.~4). This indicates unambiguously 
that these two anomalies are not artifacts due to noise but genuine features 
of phase transitions in $\alpha$-RuCl$_3$, and the possibility that the 
sample becomes degraded by applying the ultrahigh field can be excluded. 

\begin{figure*}[t]
\includegraphics[width = 1\linewidth]{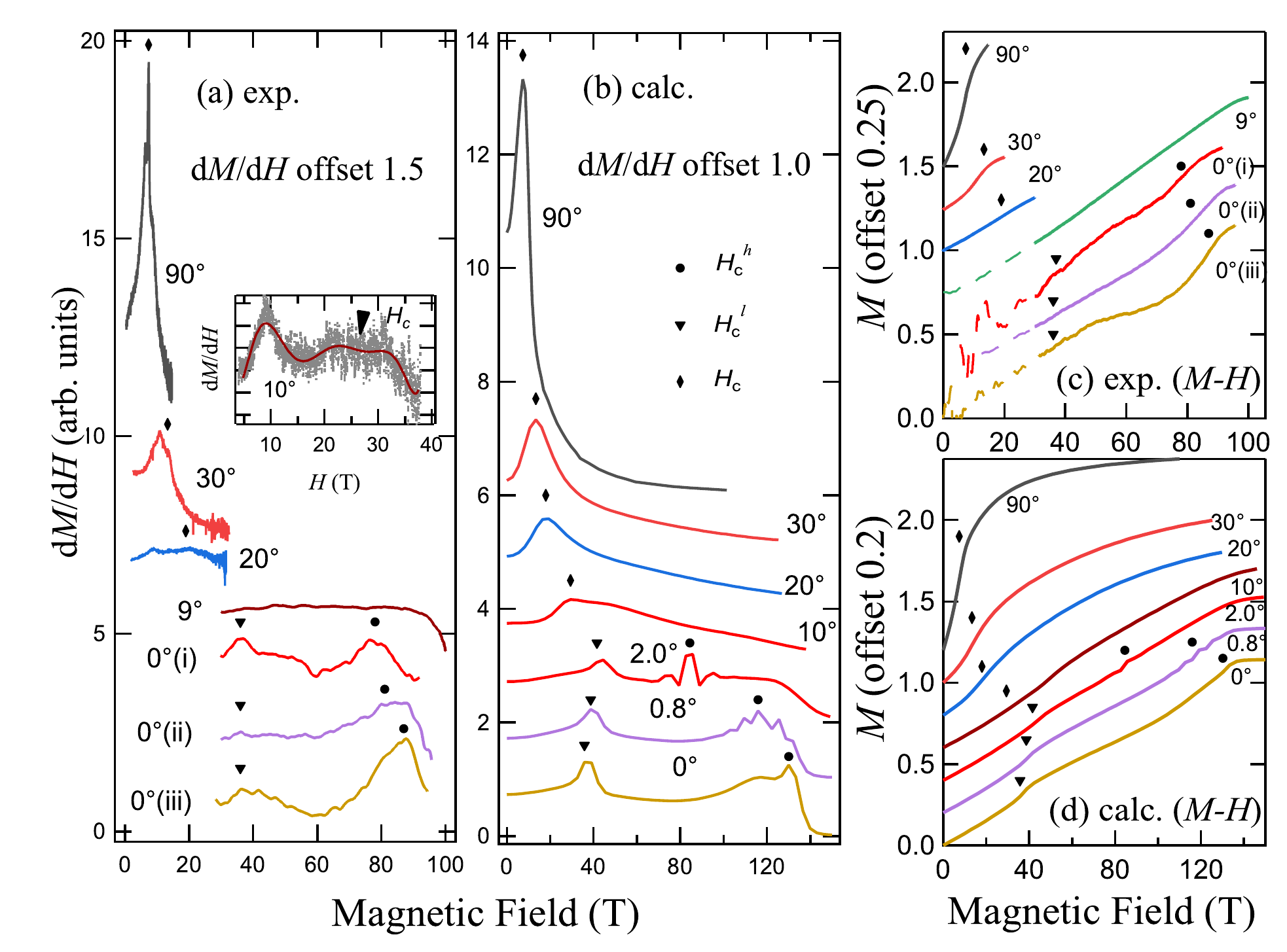}
\caption{\textbf{Comparison between the experimental 
and calculated results.} (a) The experimental d$M$/d$H$ data and 
(b) the calculated results for various $\theta$ angles.
(c) The integrated $M$-$H$ curves as well as (d) the calculated results.
The markers of diamond, triangle, and circle denotes $H_c$, 
$H_c^l$, and $H_c^h$, respectively. The experimental 
transition field at 10$^{\circ}$ are labeled in the inset of (a). Some 
calculated results in other $\theta$ angles are shown in Supplementary 
Sec.~B. The $M$-$H$ data of destructive measurements 
below 30~T are represented by dash lines.
}
\label{MHcurves}
\end{figure*}

At $\theta \simeq 9^{\circ}$ and 20$^{\circ}$, the signals in 
d$M$/d$H$ curve becomes rather weak (see also Fig.~\ref{MHcurves}) 
although we measure the data at 9$^{\circ}$ by employing the more 
sensitive pick-up coil with 1.4~mm diameter. The high-field downturn 
feature of the curve at 9$^{\circ}$ is thought to reflect the saturation 
of the magnetization as field increases. To see the transition for clarity,
we show the averaged d$M$/d$H$ curves measured by the non-destructive magnet
in the two middle insets of Fig.~\ref{exp_dmdh}, where round-peak signals 
are observed near 25 and 20 T for $\theta \simeq 9^{\circ}$ and 20$^{\circ}$. 
These round peaks in the 
middle-inset of Fig~\ref{exp_dmdh} are thought to be the phase transitions. 
The two dome structures of averaged d$M$/d$H$ curves at 9$^{\circ}$ leads to 
an uncertainty in $\theta$ of $\pm$2.5$^{\circ}$. We note that the two 
transition fields ($H_c^l$ and $H_c^h$) for $\theta \simeq 0^{\circ}$ seem 
to merge into one, and as this two curves are averaged results with $\theta 
\simeq 9 \pm 2.5^{\circ}$ and $20 \pm 2.5^{\circ}$, the peaks are very broad. 
Therefore, we define a large error bar, i.e., $\pm 5$~T for $\theta$ $\simeq$ 
9$^{\circ}$, and $ \pm 2$~T for $\theta$ $\simeq$ 20$^{\circ}$.

The results at larger angles $\theta$ $\simeq$ $30^\circ$, 
and $90^\circ$ are also shown in Fig.~\ref{exp_dmdh}~\cite{Modic2018,Modic2021}. 
The d$M$/d$H$ curve at $90^\circ$ shows two peaks and one shoulder structures. 
The peaks at 6.2~T and 7.2~T correspond respectively to the transition boundaries 
of the magnetic zigzag order (zigzag1) and another zigzag order (zigzag2), in 
agreement to previous studies~\cite{lampenkelley2021,Tanaka2022}. The shoulder 
structure seen at around 8.5~T is likely to be due to another antiferromagnetic 
(AFM) order~\cite{He2021}. Because this feature is insensitive to the field angle 
as we show in the latter part, such AFM order is deemed to be 
caused by the ABAB stacking components and the transition field is denoted 
as $H^{AB}_c$. For large angles, the critical field $H_{c}$, e.g., $H_{c}
\simeq7.2$~T for $\theta \simeq 90^{\circ}$ (i.e., in-plane), labels the upper 
boundary between the zigzag and paramagnetic phases. Such a transition 
has been widely recognized for the in-plane case as observed by neutron 
scattering experiments~\cite{Johnson2015,Banerjee2017,Banerjee2018}, and for 
tilted angles based on the magneto-torque measurements~\cite{Modic2021}. 
Besides, the additional peak at 6.2~T is generally believed to reflect 
the transition between two different zigzag antiferromagnetic phases, 
with period-3 and period-6 spin structures along the $\textbf{c}^*$ direction in the
ABC stacking, respectively, (see, e.g., Ref.~\cite{lampenkelley2021}). Here
we dub this transition field as $H^{\prime}_c$.

$H_c$ and $H^{\prime}_c$ are found to monotonously increase with decreasing 
the field angle. In contrast, $H^{AB}_c$ is independent of the field angle, 
suggesting that $H^{AB}_c$ at 8.5~T comes from a magnetically isotropic origin 
which is different from the transitions at $H_c$ and $H^{\prime}_c$. According 
to the previous study~\cite{Johnson2015,He2021}, the field location of 8.5~T 
indicates that the transition occurs in the stacking fault ABAB layers in the
sample. Therefore, the phase transition due to the suppression 
of the antiferromagnetic order in the ABAB stacking component is found to 
be isotropic, suggesting 
a 3-dimensional order which is different from the 
2-dimensional zigzag orders.

Here, we should note that the presence 
of ABAB stacking fault is almost inevitable for $\alpha$-RuCl$_3$ 
in the out-of-plane high magnetic field experiment. This is because 
the stress caused by the strong magnetic anisotropy under the
magnetic field along the $\textbf{c}^*$ axis would more or less deform 
the sample~\cite{Cao2016}. We can even damage $\alpha$-RuCl$_3$
by deforming the sample and produce lots of ABAB stacking faults, 
which now exhibits ordering temperature at about 14~K (c.f., Supplementary 
Fig.~5). Then we performed high-field experiments 
up to 100 T along the $\textbf{c}^*$ axis on this sample, where we find only 
$H_c^{\rm AB}$ peak at around $14(\pm 4)$~T. Based on the experimental
results, we conclude that the $H_c^l$ and $H_c^h$ signals should belong 
to the ABC stacking component. Furthermore, we also note that the pulse 
time of the destructive magnet is only a few microseconds~\cite{Takeyama2012},
much shorter as compared to the non-destructive magnet. This allows 
the samples to withstand less stress impulse during the measurement, 
rendering some 
advantages in measuring fragile and strong anisotropic samples such as 
$\alpha$-RuCl$_3$.

In Fig.~\ref{exp_dmdh}, by comparing the d$M$/d$H$ results at different 
$\theta$ angles from $90^{\circ}$ to $0^{\circ}$, we find strong magnetic 
anisotropy consistent with previous measurements \cite{Kubota2015,Johnson2015}.
We measured the magnetization process for $\theta \simeq 90^{\circ}$
(within the $ab$-plane) up to 90 T using the single-turn coil techniques. 
The results are shown in Fig.~\ref{exp_M}(e), which demonstrate that only 
the 7~T transition is present for $\theta \simeq 90^{\circ}$ and our 
measurements reproduce excellently the results in Ref.~\cite{Kubota2015} 
[c.f., Fig.~\ref{exp_M}~(e)]. It is found that $H_c$ monotonically 
increases with decreasing angle from $90^{\circ}$ to $0^{\circ}$, 
which is consistent with the results of Modic \textit{et al.} \cite{Modic2021}.

As we described in Fig.~\ref{exp_M}, 
the d$M$/d$H$ at $0^{\circ}$ is significantly different from 
that at large angles ($\theta \geq 9^\circ$) and exhibits 
two phase transitions. The two phase transitions indicate that an intermediate phase 
emerges between $H_c^l$ and $H_c^h$. Because Modic \textit{et al.}
\cite{Modic2021} have claimed that the zigzag order is suppressed 
for $H$ $\textgreater$ $H_c$ or $H_c^l$, the intermediate phase 
between $H_c^l$ and $H_c^h$ should be disordered and counts as 
the experimental evidence of the recently proposed QSL phase in
$\alpha$-RuCl$_3$ with fields applied along out-of-plane $\textbf{c}^*$ 
axis~\cite{Gordon2019,Lih2021}.
We also note that there is another scenario that $H_c^h$ corresponds to the transition field that
suppresses the AFM order, and $H_c^l$ just separates two different AFM phases. However, based on 
the experimental results here, the reported data of Modic \textit{et al.}~\cite{Modic2021}, and calculated results as shown in the following section, we find strong evidence that the transition at $H_c^l$ is an intrinsic characteristic of the ABC stacking component, and consider it is more reasonable
that $H_c^l$ suppresses the AFM order of the ABC stacking sample.

{\textbf{Comparison between experimental and calculated results.}}
The recently proposed realistic 
microscopic spin model with large Kitaev coupling might support our 
experimental results. We consider the $K$-$J$-$\Gamma$-$\Gamma'$ model 
$\mathcal{H}_0 = \sum_{\langle i,j \rangle_{\gamma}} [K S_i^{\gamma} S_j^{\gamma} 
+ J\,\textbf{S}_i \cdot \textbf{S}_j + \Gamma(S_i^{\alpha}S_j^{\beta} + 
S_i^{\beta}S_j^{\alpha}) + \Gamma'(S_i^{\gamma}S_j^{\alpha} + 
S_i^{\gamma}S_j^{\beta} + S_i^{\alpha}S_j^{\gamma} + S_i^{\beta}S_j^{\gamma})]$ 
($\alpha, \beta, \gamma \in \{x, y, z\}$) with parameters $K = -25$~meV, 
$J = -0.1|K|$, $\Gamma = 0.3 |K|$, and $\Gamma'=-0.02|K|$~\cite{Lih2021}. 

In Fig.~\ref{MHcurves}, we compare the experimental and calculated d$M$/d$H$ 
results as well as the integrated $M$-$H$ results. For the experimental data, 
only the critical fields associated with the pristine 
ABC stacking component, 
i.e., $H_c$, $H_c^l$, and $H_c^h$, are marked.

\begin{figure}[t]
\includegraphics[width = 0.95 \linewidth]{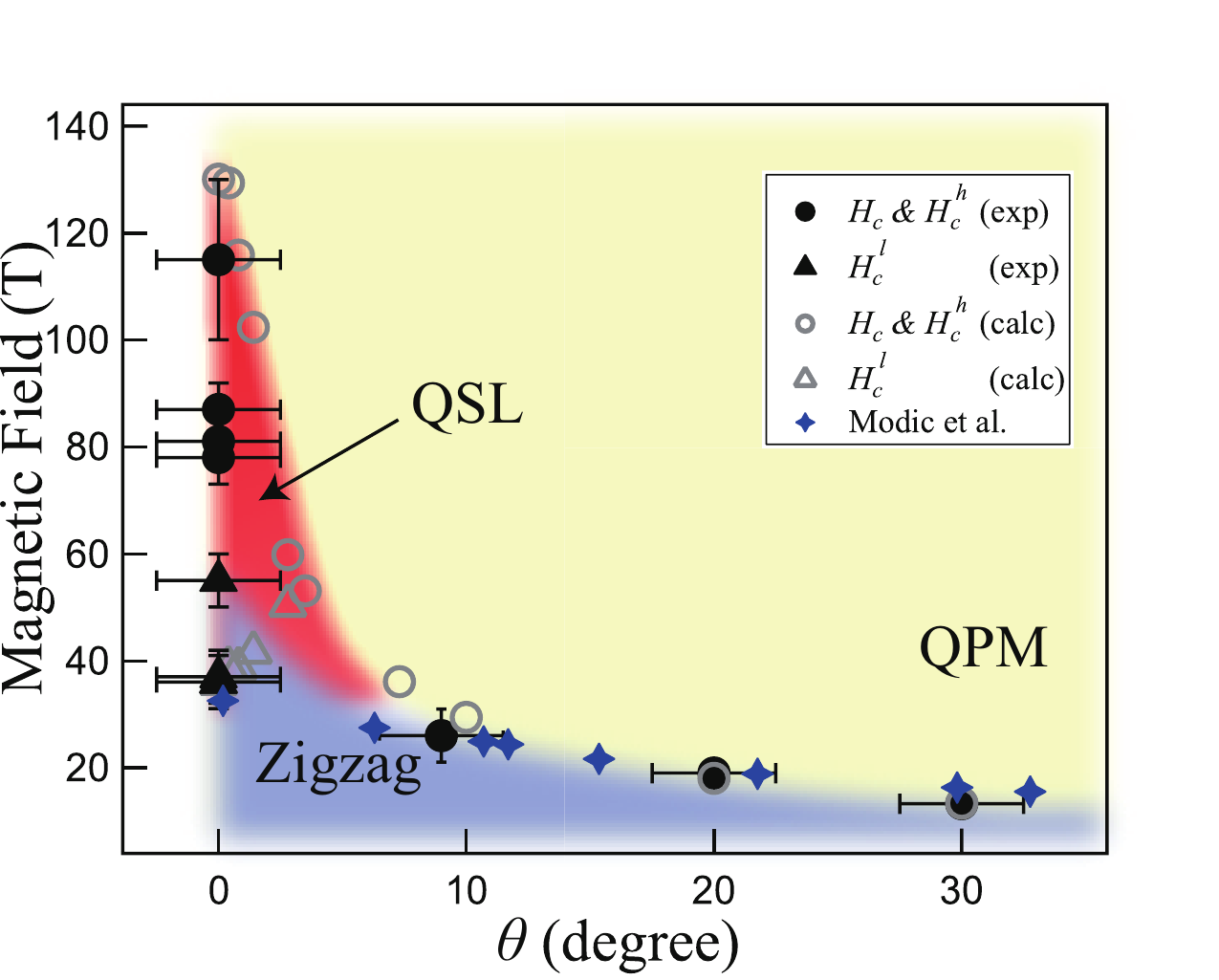}
\caption{\textbf{The field-angle phase diagram.} 
The field-angle phase diagram that summarizes the values 
of transition fields determined from both the experimental 
(black solid markers) and the calculated (grey open ones) 
$H_c, H_c^l$, and $H_c^h$. We also 
plot the low-field results (blue stars) taken from Ref.~\cite{Modic2021} 
as a supplement. The zigzag antiferromagnetic, quantum paramagnetic 
(QPM), and the quantum spin liquid (QSL) phases are indicated.
}
\label{qPD}
\end{figure}

From Figs.~\ref{MHcurves}(a,b), we find semi-quantitative agreement between 
the experimental and calculated d$M$/d$H$ results. Similarly, the experimental 
and calculated $M$-$H$ results also show consistency to each 
other as shown in Figs.~\ref{MHcurves}(c,d). In Fig.~\ref{MHcurves}(b), for 
small angles $\theta$ =0$^\circ$, 0.8$^\circ$, and 2.0$^\circ$ located within 
the angle range $\theta\simeq0^\circ \pm 2.5^\circ$, the calculated curves 
exhibit two transition fields as indicated by the solid black triangles and 
circles, and we find the upper transition fields $H_c^h$ are rather sensitive 
to the small change of $\theta$ near $0^\circ$. Therefore it explains the 
visible difference in $H_c^h$ among the three $\theta \simeq 0^\circ$ measurements. 
On the contrary, the lower transition field $H_c^l$ is found rather stable 
in Fig.~\ref{MHcurves}(b), also in agreement with experiments. As the angle 
$\theta$ further increases, e.g., $\theta = 10^\circ$, there exists a 
single transition field, in agreement with the experimental result of $9^\circ$ 
in Fig.~\ref{MHcurves}(a). The calculated transition fields $H_c$, from our
DMRG simulations based on the 2D spin model, of $\theta \simeq 20^\circ, 
30^\circ$, and $90^\circ$ cases in Fig.~\ref{MHcurves}(b) show quantitative
agreement to measurements in Fig.~\ref{MHcurves}(a). 
We note that there are still certain differences between the DMRG 
and experimental results, such as the height of peaks, which are understandable. 
The difference might be ascribed to the finite-size effects in the model 
calculations (c.f., Supplementary Fig.~7) or other possible 
terms/factors not considered in the present model study, e.g., the next- 
and third-nearest neighbor Heisenberg couplings, the inter-layer interactions, 
and the inhomogeneous external field in the high-field measurements. In 
particular, as the DMRG calculations are performed on an effective two-dimensional 
spin model, the inter-layer stacking effects in $\alpha$-RuCl$_3$ compounds
are not considered.
\\

\noindent{\bf{Discussion}}\\
From both experimental and calculated magnetization data, we see intrinsic 
angle dependence of the quantum spin states in $\alpha$-RuCl$_3$ under 
magnetic fields. Therefore, by collecting the transition fields $H_{c}^{l}$ 
and $H_{c}^{h}$ marked in Fig.~\ref{MHcurves}, we summarize the results in 
a field-angle phase diagram shown in Fig.~\ref{qPD}. In previous theoretical 
studies, an intermediate QSL phase was predicted between the upper boundary 
of zigzag phase $H_{c}^{l}$ and the lower boundary of paramagnetic phase 
$H_{c}^{h}$~\cite{Gordon2019,Lih2021}. Nevertheless, the fate of the intermediate 
QSL phase under tilted angles has not been studied before. Here we show clearly 
that the QSL states indeed constitute an extended phase in the field-angle phase 
diagram in Fig.~\ref{qPD}, as further supported by additional DMRG 
calculations of the spin structure factors here (c.f. the Supplementary Sec.~B).
Moreover, 
when $\theta$ becomes greater than about $9^{\circ}$, there exists 
only one transition field $H_c$ in Fig.~\ref{qPD}, which decreases monotonically 
as $\theta$ further increases. The previously proposed magnetic transition points 
determined by the magneto-torque measurements~\cite{Modic2021} are also plotted 
in Fig.~\ref{qPD} and found to agree with our $H_c$ for $\theta$ from $9^{\circ}$ 
to $90^{\circ}$. In addition, the two transitions ($H_c^l$ and $H_c^h$) experimentally 
obtained at $\theta \simeq 0^{\circ}$ are semiquantitatively reproduced by the 
theoretical simulation, which indicates the existence of the intermediate QSL phase. 
The transition field of the magneto-torque measurements~\cite{Modic2021} at $\theta 
= 0^{\circ}$ is also found to be in agreement with our results. For $0^{\circ} \; 
\textless \;\theta \; \lesssim  \; 9^{\circ}$, there is a discrepancy between the 
theoretical simulation and the results of the torque measurement. Although the 
reason of the difference is not completely clear 
at present, the quantum fluctuations in the vicinity of the potential tricritical 
point where the $H_c^l$ and $H_c^h$ merge disturbs the precise evaluation of 
the transition field experimentally as well as numerically. Nevertheless, the 
theoretical prediction of the extension of the QSL phase to the finite small 
$\theta$ is likely to be supported by different experimental $H_c^h$ at $\theta 
\simeq 0^{\circ}$ with $\pm 2.5^{\circ}$ uncertainty.

In summary, we find experimentally an interesting two-transition scenario 
in the prime Kitaev material $\alpha$-RuCl$_3$ under high out-of-plane 
fields up to 100-T class and reveal the existence of a field-induced 
intermediate phase in the field-angle phase diagram. Such a magnetic 
disordered phase is separated from the trivial polarized state by a 
quantum phase transition, suggesting the existence of the long-sought 
QSL phase as predicted in previous model studies~\cite{Gordon2019,Lih2021}. 
Regarding the nature of the intermediate QSL phase, previous theoretical work  
\cite{Gordon2019} concludes the intermediate QSL phase can be adiabatically 
connected to the Kitaev spin liquid (KSL) phase. On the other hand,
Ref.~\cite{Lih2021} draws a different conclusion of gapless QSL in the 
intermediate regime based on results with multiple many-body approaches. 
Here we further uncover that 
the intermediate phase also extends to a finite-angle regime, whose precise 
nature calls for further theoretical studies. While the phase diagram in 
Fig.~\ref{qPD} excludes the presence of an in-plane QSL phase like certain 
other recent studies~\cite{Bachus2020,Modic2021}, our work nevertheless 
opens the avenue for the exploration of the out-of-plane QSL phase in the 
Kitaev materials.
Moreover, further experimental characteristics of the intermediate QSL phase 
can be started from here. For example, nuclear magnetic resonance and electron 
spin resonance spectroscopy under high fields~\cite{Meier2016,
Akaki2018} are promising approaches for probing low-energy 
excitations in the intermediate QSL phase discovered here.\\

\noindent{\bf{Methods}}\\
\textbf{Experimental details.}
A single crystal of $\alpha$-RuCl$_3$ was used for the present 
experiment~\cite{Kubota2015}. The vertical-type single-turn coil 
field generator was employed to provide a pulse magnetic field 
up to 102 T. Things inside of the coil including the sample are 
generally not damaged by the generation of a magnetic field, 
although the field generation is destructive~\cite{Takeyama2012}.
The magnetization processes under the out-of-plane fields 
(Fig.~\ref{exp_M}) and those with small tilting angles ($9^\circ$ 
lines in Fig.~\ref{exp_dmdh}) were measured using a double-layer 
pick-up coil that consists of two small coils compensating for 
each other~\cite{Zhou2020,Takeyama2012}. 
The sample is cut to 0.9$\times$0.9~mm$^2$ square. Several 
sample with $\sim$0.2 mm thickness are stacking together to obtain 
enough thickness to measure the magnetization process in the 
single-turn coil experiments. The angle between the magnetic 
field and the $\textbf{c}^*$ axis is denoted as $\theta$ (c.f. upper 
inset of Fig.~\ref{exp_dmdh}). In order to have good control 
on the angle $\theta$, two glass rods with a section inclination 
angle $\theta$ are employed to clamp the sample in a Kapton tube. 
The single-turn coil, pick-up coil, and the Kapton tube with the 
sample are placed in parallel visually. As the $\alpha$-RuCl$_3$ 
sample is very soft and has strong anisotropy, it needs to be 
carefully fixed. Silicone grease instead of cryogenic glue is 
used to hold the sample, in order to reduce the dislocation of 
stacking caused by pressure (For more information of the set-up 
around the sample, see in Supplementary Fig.~3). 
Nevertheless, such an experimental setting inevitably affects 
the precise control of $\theta$ with errors estimated to be 
$\pm 2.5^{\circ}$.

Two types of double-layer pick-up coils are employed in the measurements; 
one is the standard type with 1 mm diameter~\cite{Zhou2020}, and the other is 
a recently developed one with a larger diameter of 1.4 mm that helps to enhance 
the signal by nearly three times. 
The magnetization signal is obtained by subtraction of the background signal from the sample signal, which are obtained by two successive destructive-field measurements~\cite{Takeyama2012, Matsuda2013, Zhou2020} without and with the sample (see Supplementary Fig. 3), respectively. Magnetization measurements at certain large angles like $\theta \simeq 9^\circ, 
20^\circ, 30^\circ$, and $90^\circ$ are performed by a similar induction 
method employing non-destructive pulse magnets~\cite{Kindo2001}. 
In the non-destructive pulse 
field experiment, the diameter of the sample is about 2~mm.
All of our experiments are performed at a low temperature of 4.2 K.

\textbf{Density matrix renormalization group calculation.}
We simulate the system on the cylindrical geometry up to width 6 
(c.f. Supplementary Sec.~B), and retain $D=512$ bond states that lead 
to accurate results (truncation errors less than $\epsilon \simeq 1\times 
10^{-6}$). The direction of the magnetic field $H$ is represented by 
$[l\ m\ n]$ in the spin space $(S^x, S^y, S^z)$, and the Zeeman term 
reads $\mathcal{H}_H = g \mu_B \mu_0 H_{[lmn]} 
\frac{lS^x+mS^y+nS^z}{\sqrt{l^2+m^2+n^2}}$ with $H_{[l=1,m=1,n]}$ 
tilting an angle $\theta = \arccos(\frac{2+n}{\sqrt{6+3n^2}}) 
\cdot \frac{180^{\circ}}{\pi}$ to the $\textbf{c}^*$ axis within the a$c^*$-plane, 
and the Land\'e $g$-factor is fixed as $g\simeq 2.3$. The magnetization 
curves shown in Fig.~\ref{MHcurves}(b) are obtained by computing 
$M = g \mu_B \frac{l\langle S^x\rangle+m\langle S^y \rangle + 
n\langle S^z\rangle}{\sqrt{l^2+m^2+n^2}}$.\\

\noindent{\bf{Data availability}}\\
The data that support the findings of this study are available
from the corresponding author upon reasonable request.

$\,$\\
\textbf{Acknowledgements} \\
X.-G.Z. thank Yuan Yao for fruitful discussions, 
and acknowledge Yuto Ishi, Hironobu Sawabe, and Akihiko Ikeda 
for experimental supports. 
W.L. and H.L. are indebted to Shun-Yao Yu, Shou-Shu Gong, Zheng-Xin 
Liu, and Jinsheng Wen for helpful discussions. 
X.-G.Z was supported by a JSPS fellowship.
X.-G.Z. and Y.M.H. was funded by JSPS KAKENHI No. 22F22332.
Y.H.M. was funded by JSPS KAKENHI, Grant-in-Aid for
Scientific Research (Nos.JP23H04859 and JP23H04860),  	
Grant-in-Aid for Scientific Research (B) No. 23H01117, and
ENHI Challenging Research (Pioneering) No. 20K20521. H.L. and W.L. 
were supported by the National Natural Science Foundation of China (Grant 
Nos.~12222412, 11834014, 11974036, and 12047503), CAS Project for Young 
Scientists in Basic Research (Grant No. YSBR-057), and China National 
Postdoctoral Program for Innovative Talents (Grant No. BX20220291). 

$\,$\\
\textbf{Author contributions} \\
Y.H.M and W.L supervised the project. X.-G.Z and Y.H.M performed the destructive 
magnetic field experiment. X.-G.Z, A.M and K.K performed the non-destructive field
experiment. N.K and H.T provided the sample $\alpha$-RuCl$_3$. 
X.-G.Z and Y.H.M analyzed the experimental data.
H.L, W.L and G.S performed the model calculations and analyzed the numerical results.
X.-G.Z, H.L, Y.H.M and W.L contributed to the preparation of the draft.

$\,$\\
\textbf{Competing interests} \\
The authors declare no competing interests.

$\,$\\
\textbf{Additional information} \\
Supplementary Information is available in the online
version of the paper.

\clearpage
\onecolumngrid
\setcounter{figure}{0}
\renewcommand\thefigure{S\arabic{figure}}

\begin{center}
\textbf{Supplementary Materials}
\end{center}

\date{\today}

\setcounter{section}{0}
\setcounter{figure}{0}
\setcounter{equation}{0}
\renewcommand{\thesection}{\normalsize{Supplementary Sec.~\arabic{section}}}
\renewcommand{\theequation}{S\arabic{equation}}
\renewcommand{\thefigure}{S\arabic{figure}}

\section{A.~More experimental data}

In this section, we show the magnetization results for $\theta \simeq 
90^{\circ}$, i.e., under in-plane fields. As the in-plane magnetization 
has been intensively studied, the comparison below serves as a 
benchmark of the sample as well as the precision of our measurements.
The $M$-$T$ data for the sample after the single-turn coil pulse field 
experiments are also shown in this section, which is measured by SQUID.
  
\begin{figure}[h!]
\includegraphics[width = 0.5\linewidth]{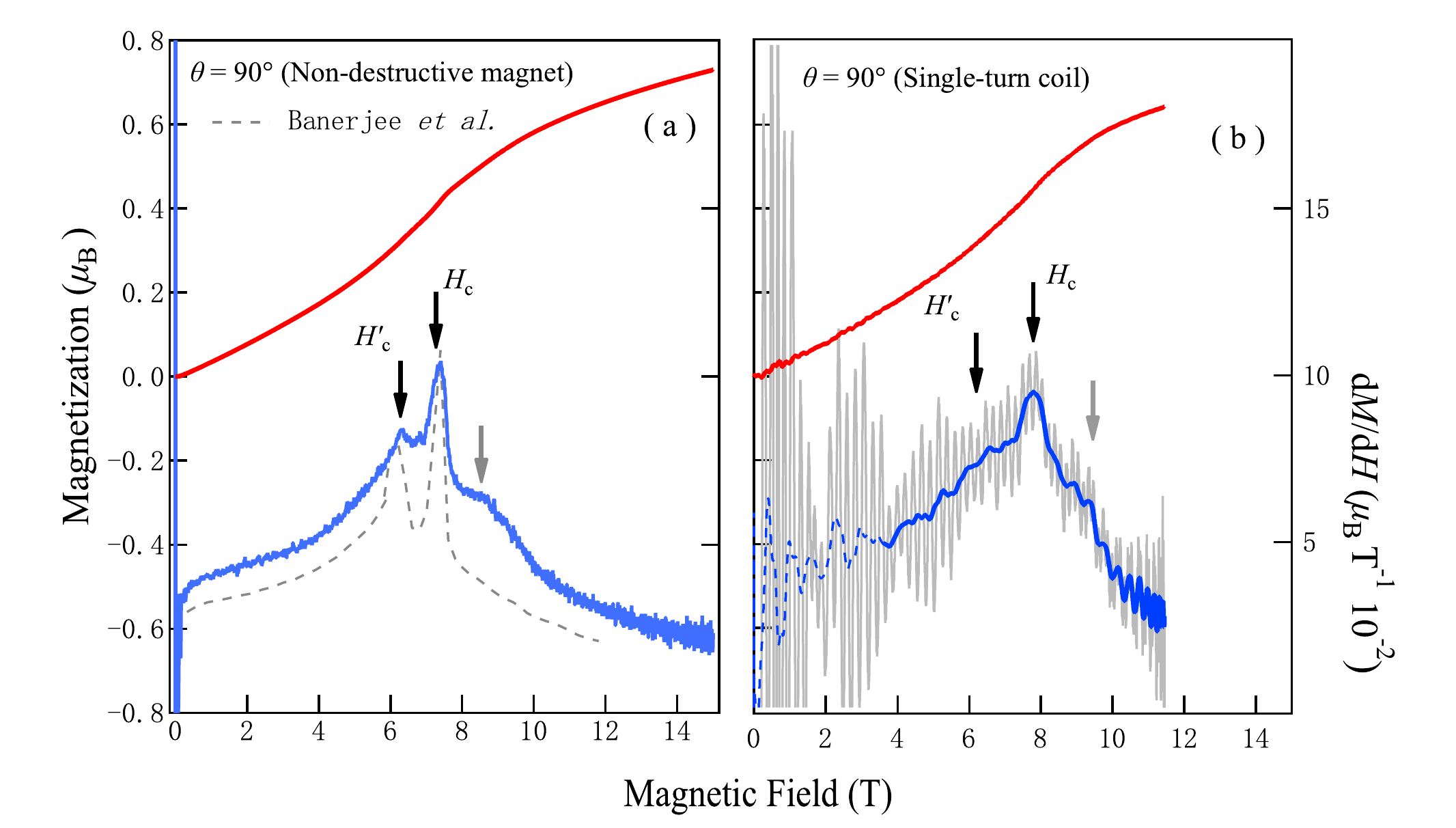}
\caption{{The magnetization curve and {d$M$/d$H$} data for 
$\theta$ = 90$^{\circ}$ measured by (a) non-destructive magnet. 
The grey dash line is the magnetic susceptibility 
results reported by Banerjee \textit{et al.}~\cite{Banerjee2018} 
(which is ac susceptibility), in which two peak 
positions (indicated by the arrows) are in very good agreements with that of ours, except for the $H_c^{AB}$
peak labelled by grey arrow.} \\
}
\label{exp_M_90}
\end{figure}
\textbf{In-plane field magnetization process.}
Supplementary Figure~\ref{exp_M_90} shows the magnetization and d$M$/d$H$ curves. 
In the {d$M$/d$H$} curve there exist three clear peaks at 6.2~T, 7.3~T, 
and 8.5~T, respectively, which are in excellent agreement with previous 
pulse-field measurements in Ref.~\cite{Kubota2015}. We have also plotted 
the magnetic susceptibility data from Ref.~\cite{Banerjee2018}, and find the 
8.5~T peak marked with the grey arrow, which can be ascribed to 
the transition in the ABAB stacking fault component, 
which is absent in their sample. The peaks at 6.2~T and 7.3~T 
are marked as $H_{c}^{\prime}$ and $H_{c}$, respectively, which have been 
proposed to be the transition fields of two different zigzag antiferromagnetic 
phases~\cite{Balz2019,lampenkelley2021}.

\begin{figure}[h!]
\includegraphics[width = 0.6\linewidth]{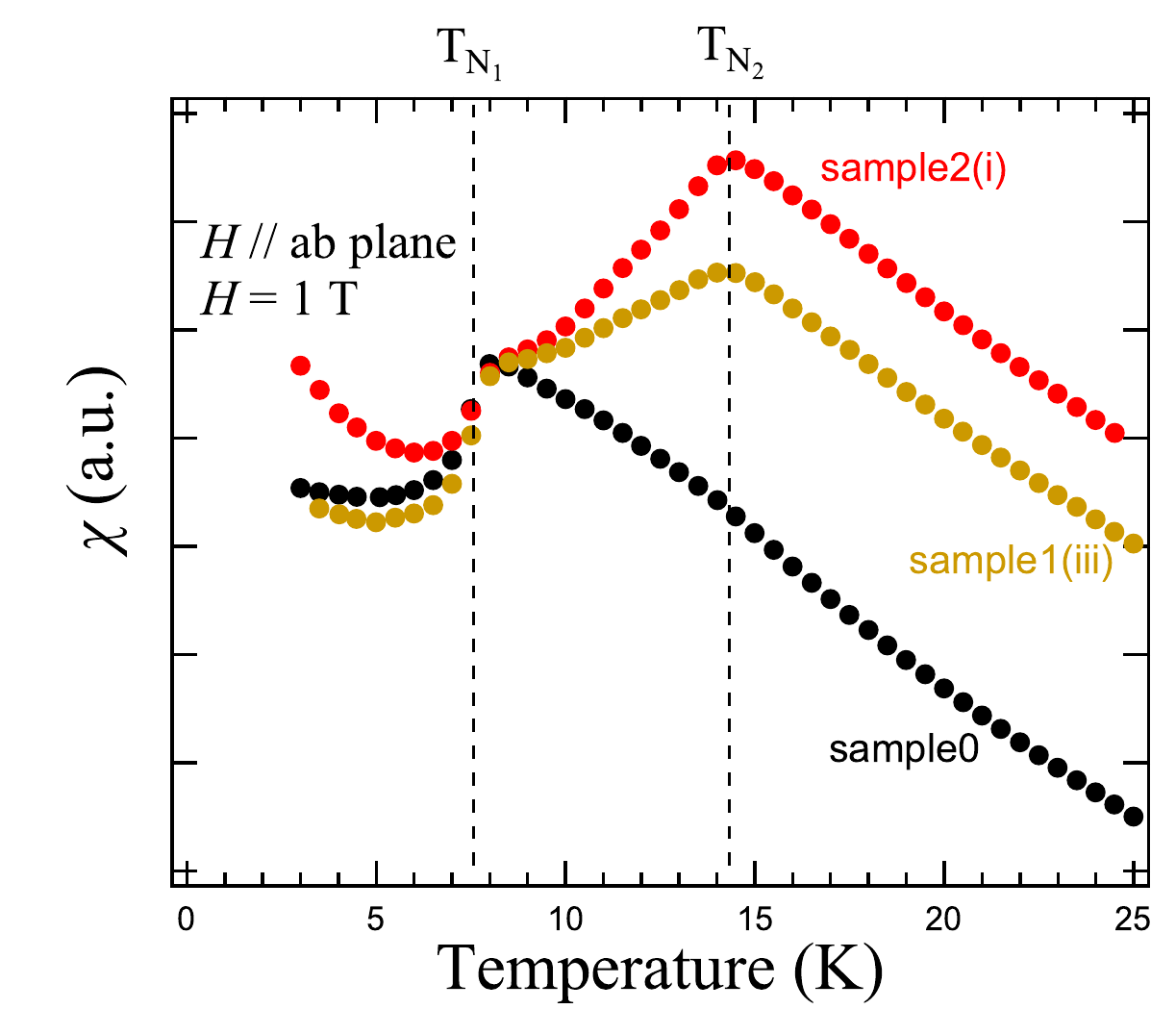}
\caption{{The $M$-$T$ curves measured by SQUID. Sample \#1  and \#2 
are measured after experiments (iii) and (i) (see Fig.~1 of the main text), respectively. 
Sample \#0 has not been exposed to single-turn coil pulse field experiments.} \\
}
\label{exp_MT}
\end{figure}

\textbf{Sample condition after high pulse field.}
The sample qualities before and after the single-turn coil pulse field are 
also checked by measuring the $M$-$T$ curves. In Fig.~1 of the main text, 
we show results of three independent single-turn coil pulse field experiments, 
namely case (i), (ii), and (iii). We have performed experiments (ii) and (iii) 
by employing the same sample (sample \#1 in Supplementary Fig.~\ref{exp_MT}), 
and case (i) by employing sample \#2. In Supplementary Fig.~\ref{exp_MT}, we have 
measured the $M$-$T$ curves of the samples after the pulse field experiments (i) 
and (iii) [as the experiment (ii) is performed before (iii)]. 
As shown in Supplementary Fig.~\ref{exp_MT}, for sample~\#1~(iii) and sample~\#2~(i) 
there are two features at 7 K and 14 K, which are ascribed to the
onset of two zigzag order with ABC- and AB-type 3D stackings,
respectively~\cite{Cao2016}, in excellent agreement with previous 
observations in Refs.~\cite{Banerjee2016, Kubota2015}. For sample \#0 
(a sample has not been exposed to the single-turn coil pulse field experiment), the 
AB-type stacking rarely appears in the sample, as evidenced by the 
very weak 14 K signal~\cite{Cao2016}. 
{In summary}, the ABC stacking is still robust in the crystal after the pulse fields expereiments.


\begin{figure}[h!]
\includegraphics[width = 0.55\linewidth]{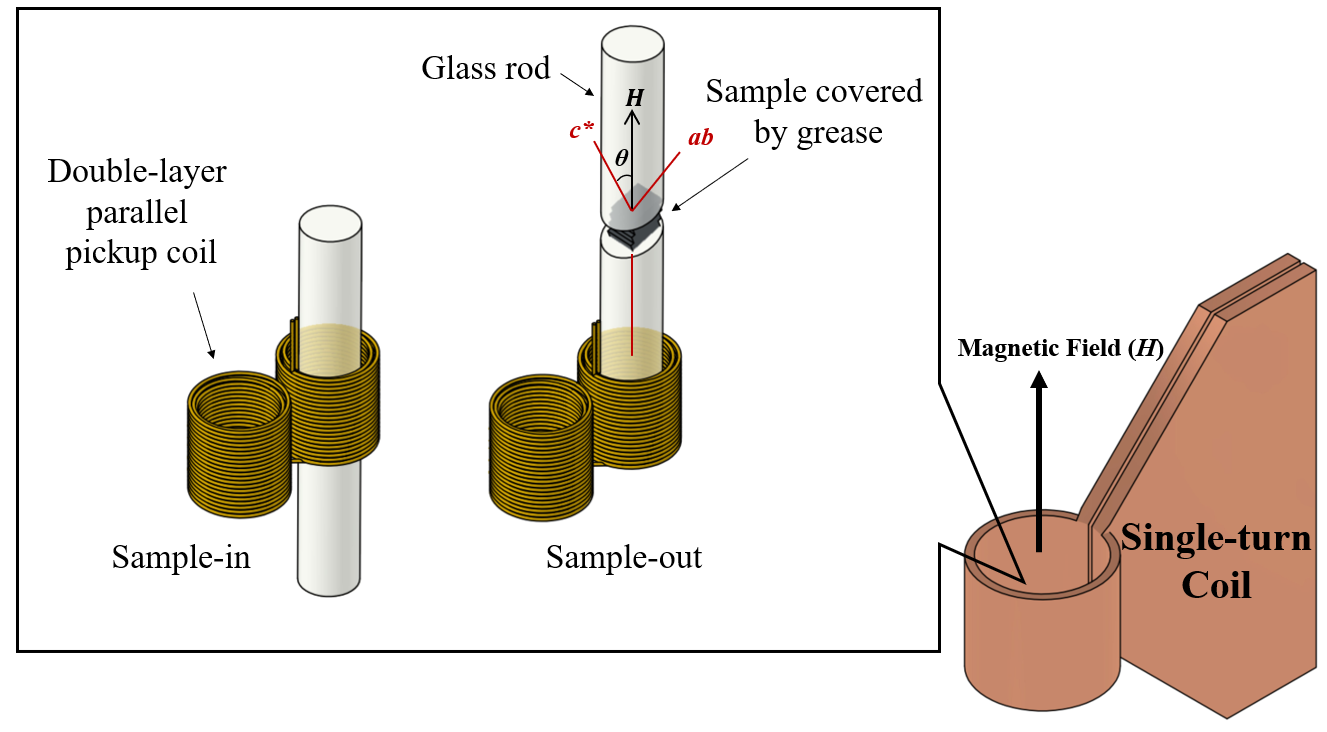}
\caption{The schematic for the setup of sample, double-layer 
magnetization pickup coil, and the single-turn coil. 
The left panel in the zoomed-in segment 
shows the condition that sample
is in the pick-up coil, while the right one shows the condition to measure the 
background of signals. \\
}
\label{FigS4}
\end{figure}

\textbf{Setup of the single-turn coil magnetic field experiment and more data.}
Supplementary Fig.~\ref{FigS4} shows the schematic for the setup of sample and double-layer 
magnetization pickup coil. When performing the single-turn coil experiment, the single-turn 
coil, pickup-coil, and the Kapton tube with sample are placed visually in parallel. The 
angle $\theta$ is determined by the inclined section the glass rod as shown in Supplementary Fig.~\ref{FigS4}.
In order to ensure that no excess stress is applied to the sample while holding the sample 
between the two glass rods, vacuum silicone grease is employed to cover the sample.
Because there are many degrees of freedom in this setup, the errorbar of angle $\theta$ between c* 
and the magnetic field is estimated to be 2.5$^{\circ}$. Some other details of single-turn 
coil magnetic field experiment has been described in Ref.~\cite{Takeyama2012, Matsuda2013, 
Zhou2020}.

In Supplementary Fig.~\ref{FigS5}, we show the comparison between the up-sweep and down-sweep 
processes of d$M$/d$H$ for experiment (i) and (iii).  In the down-sweep process
of experiment (i), we observed the signals at $\sim$8 T, $\sim$36 T, and $\sim$76~T,
which correspond to the $H_c^{AB}$, $H_c^l$, and $H_c^h$, respectively. 
The signals show excellent consistency with those observed in the up-sweep 
process. In the down-sweep data of experiment (iii), we similarly observed the 
signals at $\sim$8 T, $\sim$35 T, and $\sim$83~T (the anomalous hump at 24 T 
may be ascribed to the transition from zigzag 1 to zigzag 2). 
Here, because the signals observed at the up-sweep and down-sweep field show 
consistency with each other, and there is no significant hysteresis between the 
up-sweep and down-sweep signals, we can exclude the possibility that the second phase 
transition is caused by the pealing out of sample or the structure phase 
transition of the sample, although there are some ABAB stacking fault component increases
after our single-turn field experiment (see Supplementary Fig.~\ref{exp_MT}). In addition, the phase 
transition at $\sim$8.5 T marked by $H_c^{AB}$ is thought to be the feature of ABAB
stacking fault component in $\alpha$-RuCl$_3$, based on the Fig.~2 of our main 
text and the magnetization results in Ref.~\cite{Johnson2015, Kubota2015}.


\begin{figure}[!h]
\includegraphics[width = 0.7\linewidth]{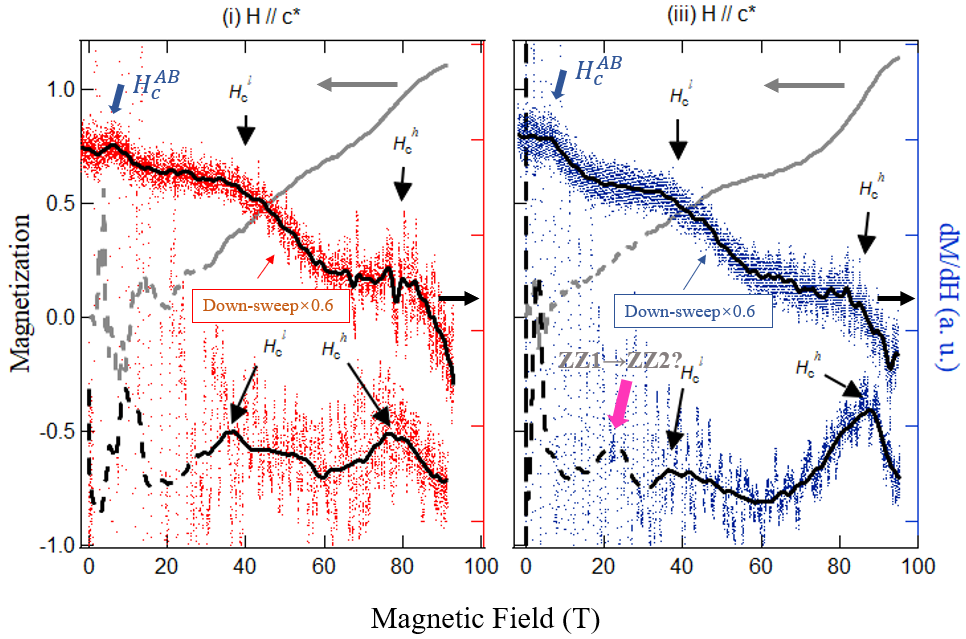}
\caption{Comparison of magnetization processes between the up-sweep 
and down-sweep d$M$/d$H$ data for experiment (i) and (iii). The gray curves 
represent the $M$-$H$ processes. The d$M$/d$H$ data are represented by 
the black curves. The anomaly marked by pink arrow may be 
ascribed to the phase 
transition from zigzag1 (ZZ1) to zigzag2 (ZZ2) which has been reported in Ref.~\cite{lampenkelley2021}.
The up-ward trend of $dM/dH$ curves at low field during demagnetization is caused 
by the inharmonious of magnetic field during down-sweep process.\\
}
\label{FigS5}
\end{figure}



\begin{figure}[h!]
\includegraphics[width = 0.5\linewidth]{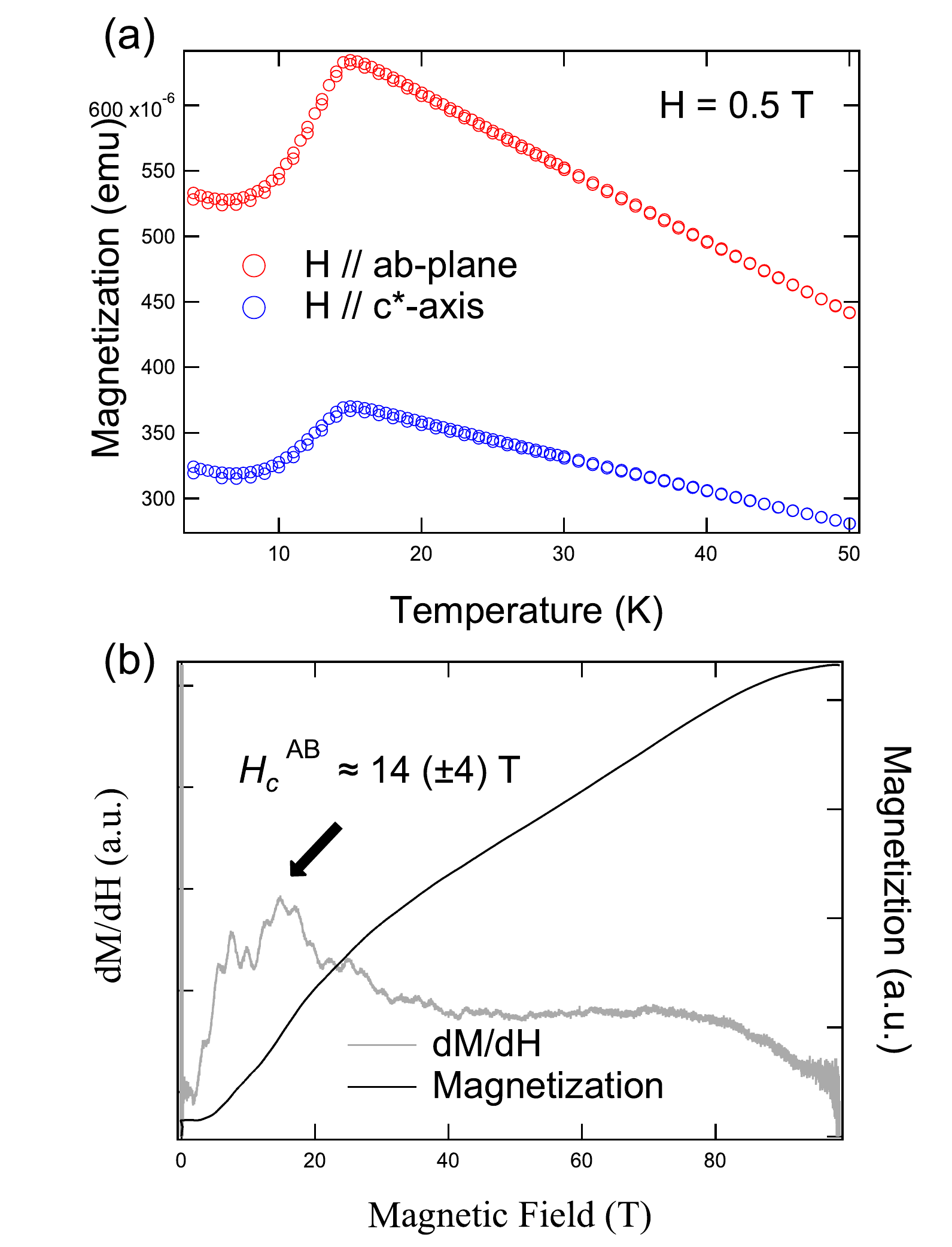}
\caption{The magnetization curve and $dM/dH$ data up to 100 T for a sample full of ABAB stacking fault components 
under the c*-axis external field. Only a 14 $\pm 4$~T peak is observed in the $dM/dH$ curves. \\
}
\label{ABstacking}
\end{figure}


To further confirm that the transition signals $H_c$, $H_c^l$, and $H_c^h$ are from the pristine
ABC stacking components but not the ABAB stacking fault, we show the magnetization 
process and $dM/dH$ data of a sample full of ABAB stacking fault under c*-axis external field up to 100 T. 
We damage $\alpha$-RuCl$_3$ by deforming the sample and producing lots of ABAB 
stacking fault components, which now exhibits ordering temperature of about 14~K, 
as shown in Supplementary Fig.~\ref{ABstacking}(a). The $dM/dH$ curve and magnetization process are given in 
Supplementary Fig.~\ref{ABstacking}(b), where the $H_c^l$ ($\sim$35~T) and $H_c^h$ ($\sim$80~T) transitions 
are absent in the ABAB stacking fault sample.
This also evidences
that the emergence of two phase transitions $H_c^l$ and $H_c^h$ in $\alpha$-RuCl$_3$ for $\theta$ is in the vicinity 
of 0$^{\circ}$, which support our conclusion of the intermediate QSL liquid phase between the 
transitions $H_c^l$ and $H_c^h$.

\section{B. Density matrix renormalization group simulations}

\begin{figure}[h!]
\centering
\includegraphics[width = 16 cm]{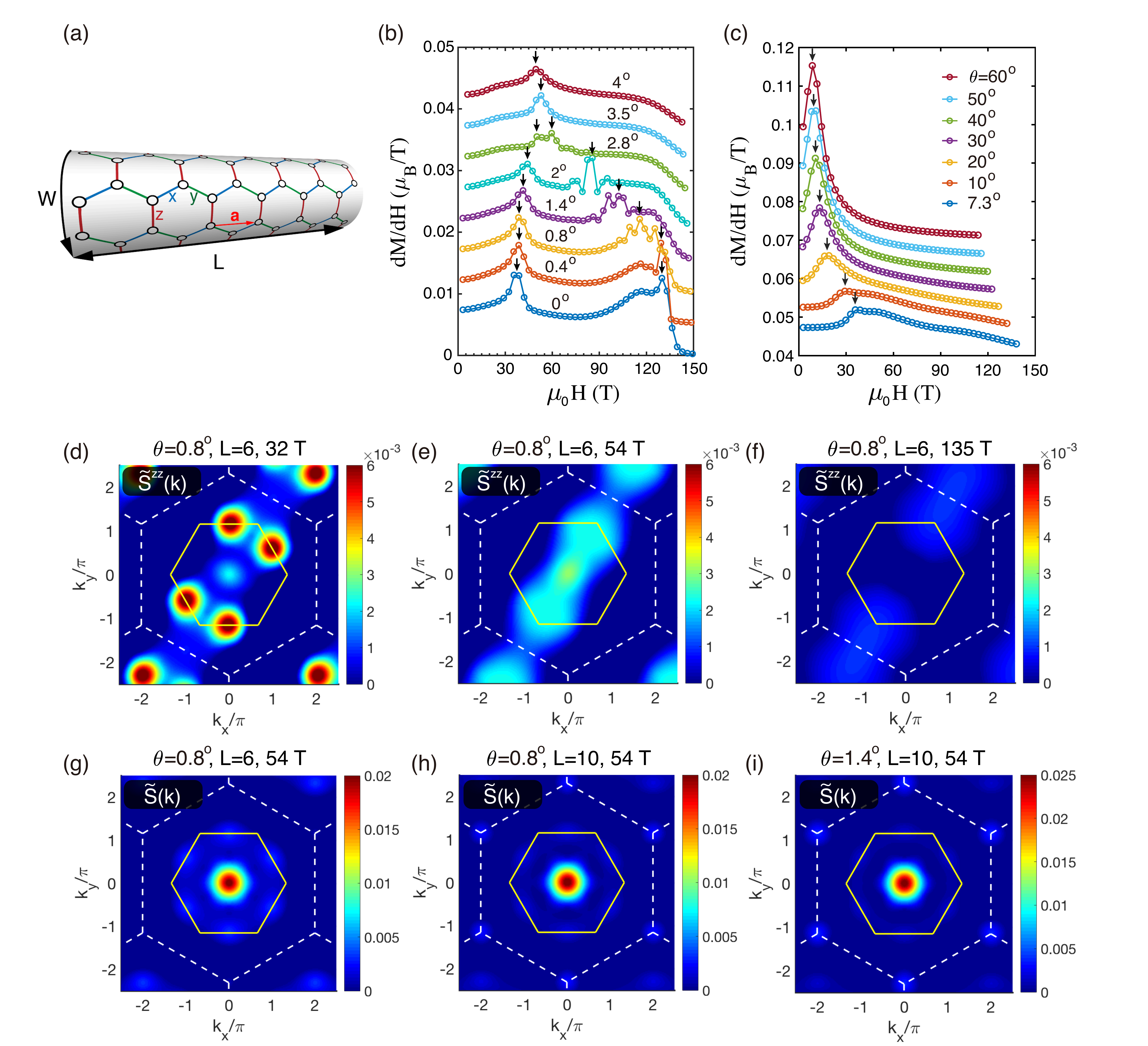}
\caption{(a) The cylindrical geometry YC$W \times L \times2$ employed 
in the DMRG simulations. (b,c) show the calculated d$M$/d$H$ curves with 
small (b) and large (c) tilting angle $\theta$. The estimated transition fields 
are indicated by the arrows. For a typical angle $\theta=0.8^{\circ}$, we show 
the contour plots of $\tilde{S}^{zz}(\textbf{k})$ under (d) $H \simeq 32$~T, 
(e) $H \simeq 54$~T and (f) $H \simeq 135$~T on YC$4 \times L \times2$ 
lattices (with $L=6$), where the stripy background as well as the peculiar 
patterns for each phase can be clearly seen. (g, h) show the total structure 
factors $\tilde{S}(\textbf{k})$ in the intermediate QSL phase, computed on 
cylinders of length $L=6$ and 10, respectively, to check the convergence 
of the results. In panel (i), we show $\tilde{S}(\textbf{k})$ with another small 
angle $\theta=1.4^{\circ}$, which resembles the results in panel (h) with $\theta=0.8^{\circ}$.}
\label{Fig:FigS3}
\end{figure}

In this section, we show the density matrix renormalization group (DMRG) 
calculations of the realistic $K$-$J$-$\Gamma$-$\Gamma^{\prime}$ 
model for $\alpha$-RuCl$_3$ under fields applied along various $\theta$ angles.\\

\textbf{The geometry adopted in DMRG simulations.}
The simulations are performed on the YC$W\times L \times2$ geometry 
with $W$ up to 6 and $L$ up to 10, with kept bond dimension $D$ up to 1024. 
The example of a YC$4\times 6 \times2$ lattice is illustrated in Supplementary Fig.~\ref{Fig:FigS3}(a),
where $x$-, $y$-, and $z$-type bond also indicated in blue, green and red 
colors. An in-plane direction [1 1 $\bar{2}$] with $\theta=90^{\circ}$ is 
indicated by the red arrow in Supplementary Fig.~\ref{Fig:FigS3}(a).\\

\textbf{The calculated d$M$/d$H$ curves and transition fields.}
As shown in Supplementary Fig.~\ref{Fig:FigS3}(b,c),
the quantum phase transition is clearly signatured by the divergent peaks 
in d$M$/d$H$. When $\theta < 3.5^{\circ}$, two {peaks at} $H_{c}^{l}$ and 
$H_{c}^{h}$ can be clearly seen, which indicate the transition fields when 
the zigzag order gets suppressed ($H_{c}^{l}$) and the system enters the 
field-induced polarized phase ($H_{c}^{h}$), respectively. For clarity, all d$M$/d$H$ 
curves are shifted vertically by about $0.1\ \mu_B /T$, and the small $\theta$ data 
are seen to suffer relatively strong finite-size effects. As $\theta$ further increases, 
we find only a single transition field $H_c$ from the zigzag-ordered phase to the 
polarized one. All the transition fields, namely, $H_{c}^{l}$, $H_{c}^{h}$ and $H_c$ 
are indicated by the arrows at the peak position of d$M$/d$H$ curves.\\

\textbf{The static spin-structure factors in various phases.}
In Supplementary Fig.~\ref{Fig:FigS3}(d-i), we show the contour plots of the spin structure 
factors in the Brillouin zone (BZ), for typical field angles $\theta=0.8^{\circ}$ 
and $1.4^{\circ}$ and computed on YC4 geometries with $L=6$ and 10. 
In particular, the $z$-component structure factors under the representative 
$H\simeq 32$~T, 54~T and 135~T are shown, 
which is described by
\begin{equation}
\tilde{S}^{zz}(\textbf{k})=\frac{1}{N}\sum_{j} 
e^{ i \textbf{k} (\textbf{r}_j-\textbf{r}_{i_o})} 
(\langle S_{i_0}^z S_j^z\rangle - \langle 
S^{z}_{i_0}\rangle\langle S^{z}_{j}\rangle),
\label{Eq:Szzq}
\end{equation}
where $N$ is the total site numbers, 
$i_0$ is a fixed central reference site,
and $j$ runs over the whole lattice.

In all three panels Supplementary Fig.~\ref{Fig:FigS3}(d-f) we see the stripy backgrounds that
represent the short-range and bond-directional spin correlation due to the strong 
Kitaev term, as discussed in Ref.~\cite{Lih2021}. The bright M points in 
Supplementary Fig.~\ref{Fig:FigS3}(d) indicates the zigzag spin correlation in the ground state, 
which gets suppressed under fields $H \geq 35$~T and the spin structure becomes 
flattened and leaves only the stripy background in Supplementary Fig.~\ref{Fig:FigS3}(e). 
As field further increases, the system enters the spin-polarized phase with 
virtually no spin correlations (Eq.~\ref{Eq:Szzq}) remained.\\

\textbf{Ground-state spin structures in the intermediate QSL phase.}
To further validate the intermediate QSL phase under small-angle fields, 
we show the contour plot of total spin structure factor $\tilde{S}(\textbf{k}) 
= \sum_{\gamma} \tilde{S}^{\gamma \gamma}(\textbf{k})$ with $\gamma
=x,y,z$ in Supplementary Fig.~\ref{Fig:FigS3}(g-i) {with $\theta=0.8^\circ$ and 1.4$^{\circ}$}. There
is no diverging peaks present in the structure factor and the brightness 
at the $\Gamma$ point remains unchanged as the system size increases 
from $L=6$ [Supplementary Fig.~\ref{Fig:FigS3}(g)] to $L=10$ [Supplementary Fig.~\ref{Fig:FigS3}(h)]. The peak at M point that
corresponds to the zigzag order gets fainter as $L$ increases, indicating 
the absence of spontaneous long-range order in the intermediate phase. 
In Supplementary Fig.~\ref{Fig:FigS3}(i), we also provide the results with slightly larger $\theta=
1.4^{\circ}$ which is very similar to Supplementary Fig.~\ref{Fig:FigS3}(h). These structure factor 
data indicate the absence of magnetic order in the intermediate-field 
regime and support the presence of a QSL phase in the field-angle 
quantum phase diagram illustrated in Fig.~4 of the main text.\\

\textbf{Finite-size effect in DMRG calculations.} 
As shown in Supplementary Fig.~\ref{Size}, in order to check the convergency of
the calculated critical fields $H_c^l$ and
$H_c^h$ in the finite-size simulations,
we show the magnetizations and their derivatives on two geometries, i.e., YC$4\times6\times2$ and YC$6\times10\times2$ lattices,
in Supplementary Fig.~\ref{Size}(a) and Supplementary Fig.~\ref{Size}(b), respectively.
It can be seen that the finite-width effect has little influence on the determination
of $H_c^l$ and $H_c^h$ for $\theta=0^{\circ}$.
In addition, we note that the height of the low-field peak at $H_c^l$ obtained by 
the larger lattice (YC6$\times$10$\times$2) 
is lower than that obtained on the small one (YC4$\times$6$\times$2), 
which seems to explain that the peaks of $H_c^l$ obtained from the experiment is relatively weaker than the calculated one in Fig.~3 of the main text.

\begin{figure}[h!]
\includegraphics[width = 0.7\linewidth]{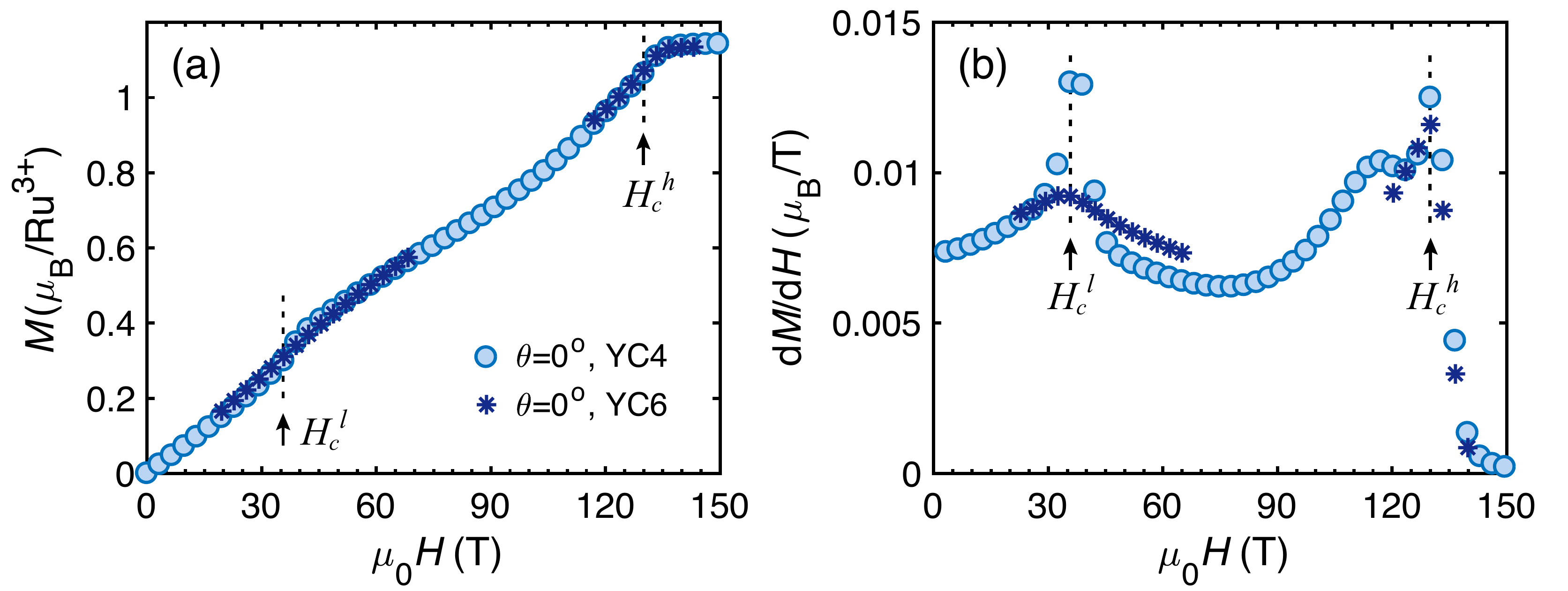}
\caption{Comparison of magnetization curves 
obtained by two finite-size DMRG calculations. 
The asterisks represent the results simulated by YC6$\times$10$\times$2, where the solid circles represent the data calculated 
on YC4$\times$6$\times$2.
The transition fields $H_c^l$ and $H_c^h$
are indicated by the black arrows, and are observed to be stable at around
35~T and 130~T for $\theta=0^{\circ}$, respectively.}
\label{Size}
\end{figure}

\sloppy

\end{document}